\def\ion#1#2{#1$\;${\small\rm\@Roman{#2}}\relax}

\documentclass[twocolumn]{aa}
\usepackage{graphicx}
\usepackage{rotating}
\usepackage{txfonts}
\begin{document}
   \title{Gas and dust properties in the afterglow spectra of GRB\,050730 
    }   

   \subtitle{}

   \author{R.~L.~C.~Starling \inst{1}, P.~M.~Vreeswijk \inst{2}, S.~L.~Ellison \inst{3}, E.~Rol \inst{4}, K.~Wiersema \inst{1}, A.~J.~Levan \inst{4,5}, N.~R.~Tanvir \inst{5}, R.~A.~M.~J.~Wijers \inst{1}, C.~Tadhunter \inst{6}, J.~R. Zaurin \inst{6}, R.~M. Gonzalez Delgado \inst{7} \& C. Kouveliotou \inst{8}
         \fnmsep
          }

   \offprints{R.~L.~C.~Starling}

   \institute{Astronomical Institute `Anton Pannekoek', University of Amsterdam,
              Kruislaan 403, 1098 SJ Amsterdam, The Netherlands.\\
              \email starling@science.uva.nl
         \and 
	 European Southern Observatory, Alonso de C\'{o}rdova 3107, Casilla 19001, Santiago 19, Chile.
	 \and
		 Dept. of Physics and Astronomy, University of Victoria, Elliott Building, 3800 Finnerty Rd, Victoria, BC, V8P 1A1, Canada.
		 \and
              Dept. of Physics and Astronomy, University of Leicester, University Road, Leicester LE1 7RH, UK.
	     	 \and
	     Centre for Astrophysics Research, University of Hertfordshire, College Lane, Hatfield, Herts. AL10 9AB, UK.     
	\and
		Dept. of Physics and Astronomy, The Hicks Building, University of Sheffield, Sheffield S3 7RH, UK.
	\and
	        Instituto de Astrofisica de Andalucia (CSIC), PO Box 3004, 18080 Granada, Spain.			
         \and
	 NASA Marshall Space Flight Center, NSSTC, XD-12, 320 Sparkman Drive, Huntsville, AL 35805, USA.}

   \date{Received ; accepted }

   \abstract{We present early WHT ISIS optical spectroscopy of the afterglow of gamma-ray burst GRB\,050730. The spectrum shows a DLA system with the highest measured hydrogen column to date: $N$(\ion{H}{I}) = $22.1 \pm 0.1$ at the third-highest GRB redshift $z$~=~3.968. Our analysis of the {\em Swift} XRT X-ray observations of the early afterglow show X-ray flares accompanied by decreasing X-ray absorption. From both the optical and the X-ray spectra we constrain the dust and gas properties of the host galaxy. We find the host to be a low metallicity galaxy, with low dust content. Much of the X-ray absorbing gas is situated close to the GRB, whilst the H~I absorption causing the DLA is most likely located further out.

   }
   \titlerunning{Gas and dust around GRB\,050730}
   \authorrunning{Starling, Vreeswijk, Ellison et al.}
   \maketitle
%
%________________________________________________________________

\section{Introduction}
Gamma-ray bursts (GRBs) have proven to be excellent probes of the distant
Universe. High luminosity GRB afterglows allow absorption line
studies of the ISM at high redshift to at least $z=4.5$ (see Andersen et
al.~2000). The launch and successful operation of the {\em Swift}
satellite means more GRBs are being localised and afterglows studied.
Subsequently, the number of high redshift bursts suitable for host galaxy spectral studies
has dramatically increased.
Deep observations of afterglow positions have detected host galaxies in
almost all cases (e.g. Conselice et al. 2005).
Most hosts are compact, actively star-forming galaxies and, where the relevant data are available, are found to have low metallicity and low intrinsic extinction (e.g. Berger et al.~2003; Tanvir et al. 2004; Christensen et al.~2004). However, 
in a few cases, radio/submm observations of hosts give a star-formation rate (SFR) 
which is of order a few to $\sim$100 times larger than rates derived from optical estimators 
such as the line luminosities of H$\alpha$ and [O~II] or the 
2800 \AA\ 
restframe UV continuum flux (e.g. Berger et al.~2003). This may be caused by strong dust
obscuration, but neither spectra nor
colours of hosts show strong internal extinction.
Afterglow spectroscopy provides a unique window on
the near environment of GRBs (e.g.~GRB\,021004, Schaefer et al. 2003; Fiore et al. 2005; Starling et al. 2005), allowing us to probe the absorbing dust and gas properties in more detail. 
In this Letter we present optical and X-ray spectra of GRB\,050730,
discovered by {\em Swift} on July 30th 2005, 19:58:23 UT
(Holland et al. 2005) and lying at a redshift of $z=3.97$ (Chen et al. \cite{GCNMIKE1}; Rol et al. 2005), in which we study the circumburst gas and dust properties.

\section{The optical afterglow spectra }
\subsection{Observations}   
During the afterglow phase of GRB\,050730, we acquired spectra using
the Intermediate-dispersion Spectroscopic and Imaging System (ISIS) on the
William Herschel Telescope. The R316R and R300B grisms were used on the
red and blue arms respectively. Two observations were done sequentially, at the parallactic angle,
with exposure times of 1260 and 1800 seconds. The first observation started at
22:57 UT at airmass $\sim$2.73 (midpoint 0.132 days after burst), 
the
second at 23:19 UT (midpoint 0.145 d) 
and airmass $\sim$3.4. The seeing quality at the high
airmasses required the slit width to be widened to 2.5 arcsec. Conditions during the observations were not photometric. These factors mean that our absolute flux calibration is not reliable, but the relative calibration should not be affected. Both
spectra have been reduced using the data reduction
package {\small IRAF} following standard procedures. 
A Galactic extinction correction of $E(B-V) = 0.049$ (Schlegel et al. \cite{Schlegel98}) was applied. The wavelength resolutions of blue- and red-arm spectra respectively are 8.7 and 8.1 \AA. The signal to noise per pixel, measured at 6800 \AA, is 27 in the first and 17 in the second spectrum.   
   \begin{figure*}
   \centering
   \includegraphics[angle=90,width=\textwidth]{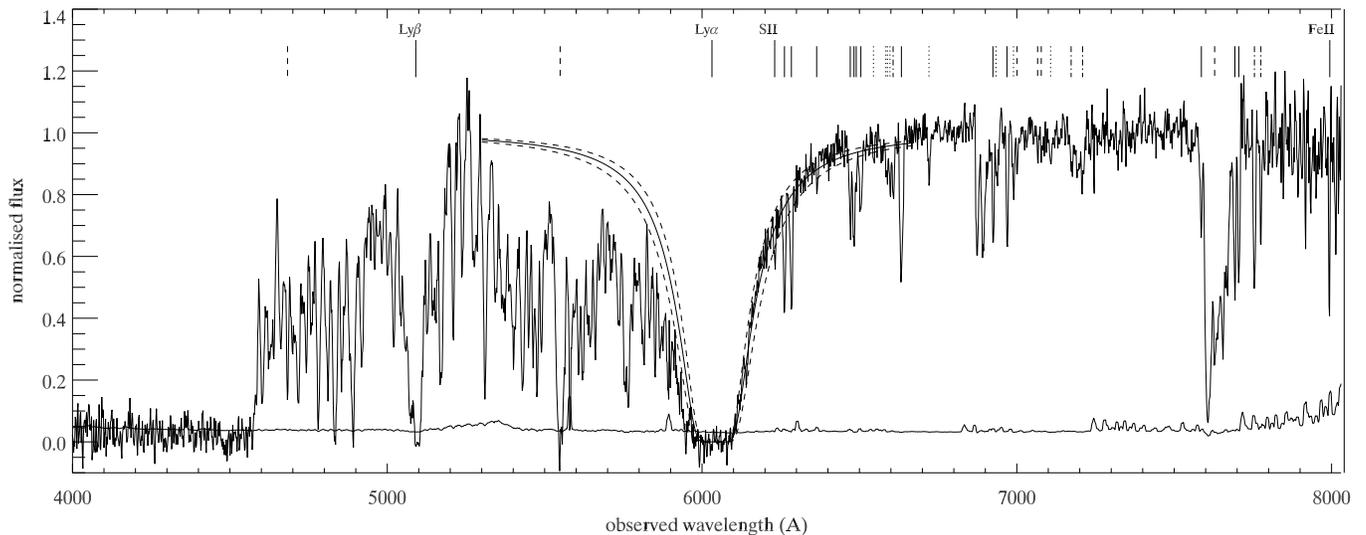}
   \caption{The WHT ISIS combined, normalised spectrum of the afterglow (midpt 0.14 days), and 1$\sigma$ error spectrum (lower curve). Overlaid is the best-fitting DLA profile (solid line) and its errors (dashed lines). All significant lines (3$\sigma$) are indicated above for $z=3.969$ (solid), $z=3.565$ (dashed), $z=1.773$ (dot-dashed) and unidentified (dotted) systems; see on-line table for details. Lines used in further analysis are labelled.
              }
         \label{spectrum}
   \end{figure*}
\subsection{Results \label{optres}}
The spectrum, shown in Fig.~\ref{spectrum}, is rich in line features at $z=3.97,3.56$ and 1.77. A strong Damped Lyman-Alpha absorption system (DLA) is present; here we focus on this and a selection of metal lines presumed to originate in the GRB host galaxy. We fitted a power law continuum corrected for Galactic extinction to the $\sim$6500--7500 \AA\ region of each spectrum, excluding the absorption lines, and find an epoch averaged slope of $\beta = -1.34\pm0.21$ (2$\sigma$ formal fit error). We tested for any departure from a pure power law due to host-galaxy extinction: fitting MW, LMC and SMC extinction curves (Pei 1992) all result in epoch averaged $A_V$ = 0.01.
The optical/IR spectral slope from published $BVRIJ$ photometry extrapolated to a common epoch using a temporal decay slope of 0.89 (Haislip et al.~\cite{GCNPROMPT}; Holman et al.~\cite{GCNHolman}; Cobb et al.~\cite{GCNSMARTS}; Blustin et al.~\cite{GCNSwiftuvot}) gives $\beta \sim -1$, consistent with the spectral analysis but not very constraining.

Despite the moderate dispersion of the ISIS 300 grisms, the damping
wings of the host galaxy DLA are clearly visible. In fact, the 
determination of $N$(\ion{H}{I}) in DLAs based on long slit spectra is
considerably simpler than for echelles. Since the damped profile
may extend over many spectral orders in a typical echelle, accurate
combination and flux calibration can be troublesome. Using the Starlink
software {\small DIPSO}, we determine log $N$(\ion{H}{I})~=~22.1~$\pm~0.1$ (see Fig.~\ref{spectrum}). 
Taking Ly$\beta$ into
account did not lead to a more accurate determination of $N$(\ion{H}{I}), and the
error on our fit is dominated by uncertainties in the determination of the power law continuum.
 The $N$(\ion{H}{I}) value is consistent with, 
although slightly lower than, that reported by Chen et al.~(2005).
This high
value (the first DLA to break the 10$^{22}$ atoms cm$^{-2}$ barrier)
continues the trend amongst GRB DLAs towards very high neutral
hydrogen columns (e.g. Jensen et al. 2001; Hjorth et al. 2003; Vreeswijk et al.~2004).

Although our spectra do not enable as detailed a study of the metal lines
as is possible via echelle observations (e.g. Chen et al. 2005), we briefly comment on a selection of these.
Detection limits are quoted at the 3 $\sigma$ level.  Although we detect
both \ion{S}{II} $\lambda, \lambda$ 1253, 1259, both lines are likely to be at
least
partially saturated. In addition, the weaker \ion{S}{II} $\lambda$ 1253 line
which
potentially offers a better limit on $N$(\ion{S}{II}) is blended with another
(unidentified) feature (H.-W. Chen, 2005, private communication). We
determine an upper limit of [S/H] $< -2.0$ based on the absence of
the weaker \ion{S}{II} $\lambda$ 1250 \AA\ line\footnote{S abundance of log (S/H)
+ 12 = 7.20 (Grevesse \& Sauval 1998)}, in good agreement with Chen et al.
(2005). Similarly, from 
the \ion{Fe}{II}~$\lambda$1608 line which is partially saturated and the
undetected \ion{Fe}{II}~$\lambda$1611, we determine $-2.9 <$~[Fe/H]~$< -1.9$.\\
A search for variability in line features between our two spectra, separated by $\sim$25 mins, revealed no significant changes (see on-line table for details), neither did the $N$(\ion{H}{I}) column vary.

\section{The absorbed X-ray afterglow \label{xrayAG}}
\subsection{Observations}
   \begin{figure}
   \centering
   \includegraphics[width=8cm]{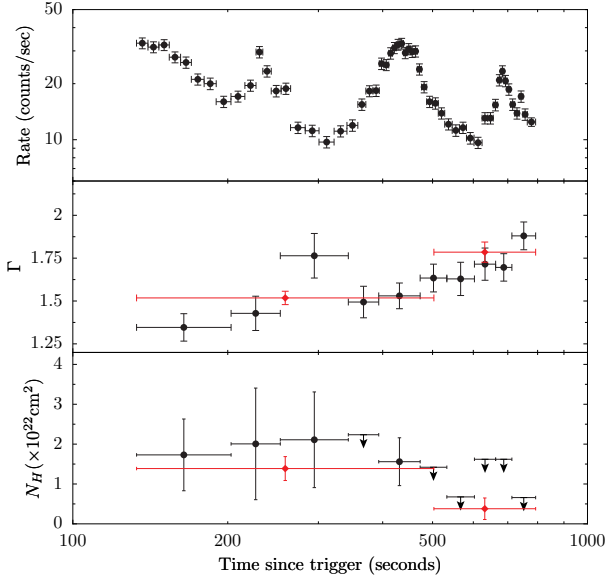}
      \caption{Evolution of the 0.3--10 keV count rate (top), 0.2--10 keV power law slope, $\Gamma$, (middle, note that $\Gamma = 1 - \beta$) and additional equivalent hydrogen column, $N_{\rm H,int}$, at the redshift of the host galaxy (bottom) during the first $\sim$800 s of the {\it Swift} XRT observations (90\% errors). The individual spectra are shown with filled circles; the two combined spectra pre- and post-500 s, for which better constraints are obtained, are shown with diamonds.
              }
         \label{xraycolumn}
   \end{figure}
We have analysed the early {\em Swift} XRT data, to look for evidence of intrinsic absorption in the X-ray spectrum. 
The XRT data consist of Windowed Timing (WT) mode data for the first orbit (133 to 793 seconds after the trigger)
 and start of the second orbit, and Photon Counting (PC) mode data for later orbits. 
The data were reduced
using the standard pipeline for XRT data within the 
{\small HEADAS 6.0} package ({\em Swift} software version 2.0).
WT mode data was extracted using a rectangular region centred on the source, and a similar area in a source-free region of the same image
to determine the background level. 
PC mode data was extracted using a circular aperture, 
except for orbits 2 to 4 which show evidence of pile-up (count rates $\gtrsim 0.8$ counts s$^{-1}$) and were extracted using an annular region centred on the source and filtered on grade 0 only.
The light curve was obtained between channels 30 and 1000 (spanning $\sim$0.3--10 keV). Spectral analysis was done using {\small XSpec 11.3}, with the standard Ancillary Response Function (ARF) files, which estimate the effective telescope area, for PC mode data, and with ARFs
based on ray-tracing (`physical' ARFs) for WT mode data which should provide a better calibration at low energies.
\subsection{Results}
The first orbit shows several flares in the light curve, first reported for this afterglow by Grupe et al. (2005). We have performed a detailed analysis of the spectral evolution of the early-time data. The fitted model consists of a power law plus Galactic absorption (fixed at $3.05 \times 10^{20}$ cm$^{-2}$, Dickey \& Lockman \cite{dickey}) and a variable Galactic-like absorption component with Solar metallicity and $z=3.97$. Errors are quoted at the 90\% confidence level for 1 interesting parameter.
We find
evidence for a change in power law photon index, from $\Gamma=1.52\pm0.04$ at the start of the first orbit to $1.79\pm0.06$ at the end of the 
orbit (note that $\Gamma = 1 - \beta$). We also find evidence for an excess absorption column, which at the redshift of the burst amounts to an intrinsic 
column of $N_{\rm H,int} = (1.4 \pm 0.3) \times 10^{22}$ cm$^{-2}$. However, around 500 seconds post trigger, the absorption column 
abruptly changes, becoming lower by about a factor of 4: $N_{\rm H,int} = (3.4 \pm 2.7)\times 10^{21}\, \mathrm{cm}^{-2}$. In the late-time PC mode spectrum the intrinsic column cannot be constrained, setting an upper limit of only $N_{\rm H,int} \le 1.0\times 10^{22}$~cm$^{-2}$, and the power law photon index remains stable at $\Gamma \sim 1.77$.  
We have checked for a possible correlation between the intrinsic $N_{\rm H,int}$
and $\Gamma$ in the fit. Contour plots for the intervals 133--503~s and 503--793~s post
trigger show no evidence for any correlation, confirming the reality of both
the drop in $N_{\rm H,int}$ and increase in $\Gamma$ (Fig. \ref{xraycolumn}).
Interestingly, 
this happens directly after the peak of the second visible flare, where the light curve intensity has increased by a factor 
of 3. 
Given the host galaxy metallicity we measure in the optical spectrum, we adjust the X-ray absorption model accordingly. Using $Z = Z_{\odot}/100$ for all the elements heavier than He included in the zvphabs X-ray absorption model, the required intrinsic equivalent hydrogen column increases by a factor of $\sim$10 in both cases to $N_{\rm H,int} = 9.5^{+2.3}_{-2.1}\times 10^{22} \mathrm{cm}^{-2}$ (first 400 s) and $N_{\rm H,int} = 2.6^{+1.9}_{-1.6}\times 10^{22} \mathrm{cm}^{-2}$ ($\ge$ 500 s post trigger) with approximately the same goodness of fit.

A preliminary analysis of published optical photometry together with the PC mode XRT spectrum has shown the 
X-ray and optical slopes at 0.19 days ($\beta_{\rm X}\sim-0.7$ to $-0.8, \beta_{\rm opt}\sim-1.0$ to -1.5) to likely be incompatible with a position of the cooling break between
the optical and xrays, and might suggest the presence of an inverse Compton component; we await the availability of further optical/IR photometry for a full analysis.

\section{Discussion and conclusions}
\subsection{Host galaxy properties}
There is a well known relationship between galaxy luminosity and
metallicity (e.g. Garnett 2002; Lamareille et al. 2004) 
which spans 6 orders or more of magnitude in $M_B$.
Tremonti et al.~(2004) have recently demonstrated that this
relation is driven by an underlying relation between mass and
metallicity. The cause of the relationship, they argue, is due
to the increased gravitational potential of massive galaxies which
enhances metal retention. In the absence
of a detected host for GRB\,050730 at the time of writing, it is in principle possible to
use the luminosity-metallicity (LZ) and mass-metallicity (MZ)
relations to predict the $M_B$ and stellar mass of the host.
Both of these relations are best determined locally (e.g. Lamareille
et al.~2004), although
sizeable datasets have now investigated the LZ relation up
to $z \sim1$ (e.g. Kobulnicky et al.~2003; Kobulnicky \& 
Kewley 2004). There is clear evidence for evolution in the
LZ relation, in the sense that galaxies are more metal-poor
for their luminosity at higher $z$ (although see caveats in
Kewley \& Ellison~in prep.). This trend appears to continue both for
the LZ and MZ relations up to 
$z \sim 3$ (e.g. Shapley et al.~2004; M\o ller et al. 2004; Erb et al.~in~prep.), 
although only the highest mass/luminosity galaxies are bright enough
to be included in spectroscopic samples. The lowest metallicity bin
in the fitted MZ relation of Erb et al.~(in prep.) is $Z \sim Z_{\odot}/3$
corresponding to a stellar mass log($M_{\star}/M_{\odot}) \sim 9.5 $.
The metallicity measured from absorption lines in the optical afterglow considered
here is $Z \sim Z_{\odot}/100$,
which indicates that the host is not a massive, luminous Lyman break galaxy (LBG), 
although Jakobsson et al. (2005) argue that GRB hosts follow the same
UV luminosity function as the faint LBGs. 
We do note that the MZ relation is based on {\em emission} lines.
However, HST imaging has shown that GRBs occur in regions of strongest star formation (e.g. Fruchter et al. 2005), 
justifying our assumption that the absorption lines are formed in the same regions 
as the higher wavelength emission lines.
Combining the measured $N$(\ion{H}{I}) with the metallicity and assuming an SMC gas-to-reddening ratio (Bouchet et al.~1985), we can estimate
the extinction associated with the GRB host galaxy and compare this to the values obtained from the optical continuum fits. 
Negligible $E(B-V)$ is determined by both methods, consistent with the similarly small amounts of dust seen
towards intervening DLAs (e.g. Murphy \& Liske 2004; Ellison, Hall \& Lira 2005).

\subsection{The neutral hydrogen column}
GRB\,050730 has the strongest DLA seen in a GRB afterglow spectrum,
with a hydrogen column density of $\log N$(H~I) = 22.1 $\pm$ 0.1.
The X-ray absorption at late times scaled to $Z_{\odot}/100$ yields a comparable
$\log{N_{\rm H,int}}=22.4^{+0.2}_{-0.4}$ (assuming the $N_{\rm H,int}$ measured at $\sim$500--800 s post burst in the WT mode XRT spectrum can be extrapolated to a few hours post burst).
The $N_{\rm H,int}$ we measure in the early-time X-ray spectra covering $\sim$133--500 s post trigger is about ten times higher than that measured at $t >$ 500 s. 
The change in X-ray absorbing column could be caused by ionisation by the gamma-ray jet, or by the X-ray flares which are suggested to be caused by
prolonged central engine activity (Burrows et al. 2005; King et al.
2005).

It should be noted that what is measured in the X-ray models is an equivalent hydrogen column, since primarily metal edges contribute to the X-ray absorption at the redshift of GRB\,050730, and that this is highly dependent upon the metallicity assumed
(see e.g. Wilms et al. 2000). There will be a contribution to the X-ray absorption from intervening systems, which cannot be disentangled from absorption in the host, particularly given that we do not know the metallicity of the closest intervening system observed in this spectrum ($z$ = 1.77). In principle, a lower column very close to the observer could have a similar effect on the spectrum as a large column at high redshift. 

The observed X-ray column variability does, however, lead us to conclude that most of
the X-ray absorbing gas in GRB\,050730 is located close to the GRB. The optical H~I column remained stable over the $\sim$25 mins between our ISIS spectra, taken at 0.132 days since burst, well after the observed X-ray flaring (although the occurrence of X-ray flares at later times cannot be ruled out owing to low count rates). The H~I creating the DLA is likely to be located much further away from the GRB, unaffected by the GRB radiation.
We would expect to observe destruction by the GRB of dust co-located with the X-ray absorbing gas. Our spectra imply a very low extinction in the host at $\sim$3 hrs post burst. Future prompt optical spectra, in conjunction with X-ray observations, are required to investigate this further. 
\begin{table}[tbp]
  \centering 
  \caption[]{Lines detected above 3$\sigma$ in the first and second
    epoch WHT spectra.$^{\mathrm{a}}$. Submitted as on-line table only. \label{tab:lines}}
    \null\vspace{-1.0cm}
  $$
    \begin{array}{lccrc}
    \hline
    \noalign{\smallskip}
    \rm \lambda  &
    W_{\rm obs}\rm(I) &
    W_{\rm obs}\rm(II) &
    \rm ID &
    z \\
    \hline
4683.0^{\mathrm{b}} &  &    &   \rm Ly\beta\,\lambda 1025.722 & 3.5656 \\
5090.6^{\mathrm{b}} &  &    &   \rm Ly\beta\,\lambda 1025.722 & 3.9629 \\
5549.2^{\mathrm{b}} &  &    &  \rm Ly\alpha\,\lambda 1215.668 & 3.5647 \\
6031.5^{\mathrm{b}} &  &    &  \rm Ly\alpha\,\lambda 1215.668 & 3.9615 \\
6230.8 & 1.42 \pm 0.24 & 1.19 \pm 0.38 &   \ion{S}{II}\,\lambda 1253.805 & 3.9695 \\
6261.9 & 3.64 \pm 0.25 & 3.50 \pm 0.41 &  \ion{Si}{II}\,\lambda 1260.422 & 3.9681 \\
       &&	       \rm blended\,with & \ion{S}{II}\,\lambda 1259.518 & 3.9717 \\
6283.9 & 3.95 \pm 0.23 & 3.45 \pm 0.33 &\ion{Si}{II}^*\,\lambda 1264.738 & 3.9685 \\
6364.2 & 0.51 \pm 0.17 & 0.78 \pm 0.35 &   \ion{C}{I}?\,\lambda 1280.135 & 3.9715 \\
6470.8 & 1.95 \pm 0.17 & 2.43 \pm 0.39 &    \ion{O}{I}\,\lambda 1302.169 & 3.9692 \\
6482.8 & 2.57 \pm 0.14 & 2.87 \pm 0.28  & \ion{O}{I}^*\,\lambda 1304.858 & 3.9682 \\
       &&              \rm blended\,with &\ion{Si}{II}\,\lambda 1304.370 & 3.9700 \\
6489.8 & 0.50 \pm 0.11 & 0.82 \pm 0.20&\ion{O}{I}^{**}\,\lambda 1306.029 & 3.9691 \\
6504.6 & 1.86 \pm 0.16 & 2.17 \pm 0.35 &\ion{Si}{II}^*\,\lambda 1309.276 & 3.9681 \\
       &&             \rm blended\,with?& \ion{Fe}{II}\,\lambda 2344.214 & 1.7747 \\
6544.6 & 0.44 \pm 0.13 & 0.47 \pm 0.17 &                                 &        \\
6583.9 & 0.42 \pm 0.10 & 0.55 \pm 0.18 &                                 &        \\
6588.9 & 0.55 \pm 0.11 & 0.57 \pm 0.17 &                                 &        \\
6597.1 & 0.82 \pm 0.12 & 1.33 \pm 0.26 &                                 &        \\
6607.0 & 1.24 \pm 0.14 & 1.58 \pm 0.23 &  \ion{Fe}{II}\,\lambda 2382.765 & 1.7728 \\
6633.4 & 5.12 \pm 0.19 & 5.07 \pm 0.31 &   \ion{C}{II}\,\lambda 1334.532 & 3.9706 \\
                  && \rm blended\,with &  \ion{C}{II}^*\,\lambda 1335.663 & 3.9664 \\
6721.0 & 1.05 \pm 0.17 & 0.73 \pm 0.20 &                                 &        \\
6924.1 & 2.43 \pm 0.15 & 2.42 \pm 0.32 &  \ion{Si}{IV}\,\lambda 1393.760 & 3.9679 \\
6934.0 & 0.88 \pm 0.09 & 1.18 \pm 0.24 &                                 &        \\
6968.7 & 2.67 \pm 0.15 & 2.32 \pm 0.28 &  \ion{Si}{IV}\,\lambda 1402.773 & 3.9678 \\
                  &&   \rm blended\,with &\ion{Si}{II}\,\lambda 1526.707 & 3.5645 \\
6989.4 & 1.44 \pm 0.17 & 1.28 \pm 0.22 &                                 &        \\
7000.9 & 0.52 \pm 0.09 & 0.86 \pm 0.24 & \ion{Si}{II}^*\,\lambda 1533.432 & 3.5655 \\
7066.1 & 0.42 \pm 0.12 & 0.20 \pm 0.19 &   \ion{C}{IV}\,\lambda 1548.204 & 3.5641 \\
7077.7 & 0.32 \pm 0.12 & 0.48 \pm 0.18 &   \ion{C}{IV}\,\lambda 1550.781 & 3.5640 \\
7107.4 & 0.62 \pm 0.14 & 0.61 \pm 0.22 &                                 &        \\
7172.6 & 0.71 \pm 0.14 & 0.92 \pm 0.25 &  \ion{Fe}{II}\,\lambda 2586.650 & 1.7729 \\
7209.0 & 1.46 \pm 0.17 & 0.86 \pm 0.20 &  \ion{Fe}{II}\,\lambda 2600.173 & 1.7725 \\
7586.4^{\mathrm{c}} & 2.53 \pm 0.18 & 2.48 \pm 0.26 &  \ion{Si}{II}\,\lambda 1526.707 & 3.9691 \\
7628.5^{\mathrm{c}} &        &          &  \ion{Al}{II}\,\lambda 1670.789 & 3.5658 \\
7692.9 & 3.79 \pm 0.14 & 3.86 \pm 0.29 &   \ion{C}{IV}\,\lambda 1548.204 & 3.9689 \\
7705.3 & 3.66 \pm 0.20 & 4.22 \pm 0.36 &   \ion{C}{IV}\,\lambda 1550.781 & 3.9687 \\
7754.9 & 4.18 \pm 0.34 & 5.15 \pm 0.63 &  \ion{Mg}{II}\,\lambda 2798.743 & 1.7709 \\
7775.5 & 2.28 \pm 0.25 & 2.67 \pm 0.36 &  \ion{Mg}{II}\,\lambda 2803.532 & 1.7735 \\
%7994.3 & 3.14 \pm 0.54 & 3.27 \pm 0.74 &  \ion{Fe}{II}\,\lambda 1611.200 & 3.9617 \\
7994.3 & 3.14 \pm 0.54 & 3.27 \pm 0.74 &  \ion{Fe}{II}\,\lambda 1608.451 & 3.9716 \\
  \end{array}
  $$
  \begin{list}{}{} 
    
    \item[$^{\mathrm{a}}$] Blueward of Ly$\alpha$ the low resolution
    and Ly$\alpha$ forest hampers secure identification of metal
    lines, which we therefore do not list.

    \item[$^{\mathrm{b}}$] Due to the uncertain continuum level, we do
    not attempt to measure the widths of the Ly$\alpha$ and Ly$\beta$
    lines.

    \item[$^{\mathrm{c}}$] This equivalent width measurement is
    seriously affected or made impossible by the atmospheric
    absorption band from 7584--7675\AA.

  \end{list}
\end{table}
\begin{acknowledgements}
We thank J.~P.~U. Fynbo for comments on and improvements to the manuscript and P.~A. Curran, A.~J. van der Horst, K.~L. Page and S. Vaughan for useful discussions.
This work is based on observations made with the WHT operated
    on the island of La Palma by the Isaac Newton Group in the Spanish
    Observatorio del Roque de los Muchachos at the Instituto de
    Astrofisica de Canarias - we thank N. O'Mahony for excellent support.
The authors acknowledge support from and collaboration within the EU-funded Research Training Network `Gamma-Ray Bursts: an enigma and a tool' (HPRN-CT-2002-00294). 
\end{acknowledgements}


\begin{thebibliography}{}

   \bibitem[2000]{GRB000131_no_DLA} Andersen, M.~I., Hjorth, J., Pedersen, H., et al. 2000
    A\&A, 364, L54 % No DLA, HG spectrum      
  
   
   \bibitem[2003]{BergerradioSFR} Berger, E., Cowie, L. L., Kulkarni, S. R., et al. 2003,
    ApJ, 588, 99
  
%   \bibitem[1996]{xspec} Arnaud, K.~A. 1996, 
%   Astronomical Data Analysis Software and Systems V, eds. Jacoby, G. \& Barnes, J., ASP Conf. Series, 101, 17
   
%   \bibitem[2005]{GRB050603_no_DLA} Berger, E. \& Becker, G. 2005,
%    GCN Circ. 3520 % No DLA, but Ly alpha in emission.    

%   \bibitem[2005]{GRB050505_DLA} Berger, E., Bradley Cenko, S., Steidel, C. et al, 2005
%    GCN Circ. 3368 % DLA, unknown strength

%   \bibitem[2003]{BergerSFR} Berger, E., Cowie, L.~L., Kulkarni, S.~R. et al. 2003,
%       ApJ, 588, 99
% 
%   \bibitem[2004]{Courty} Courty,~S., Bj\"ornsson, G., Gudmundsson, E.~H. 2004,
%       MNRAS, 354, 581
    
%GCNs on this burst to 3rd Aug, Klaas
   \bibitem[2005]{GCNSwiftuvot} Blustin, A., Holland, S.~T., Cucchiara, A., et al. 2005, 
       GCN Circ. 3717 %Swift UVOT data
     
  \bibitem[1985]{bou85}
        Bouchet, P., Lequeux, J., Maurice, E., et al. 1985, 
	A\&A, 149, 330
  
%   \bibitem[2005]{GCNRTT} Burenin, R., Tkachenko, A., Pavlinsky, M., et al. 2005,
%       GCN Circ. 3718  % R band point

   \bibitem[2003]{Burrows} Burrows, D. N., Romano, P., Falcone, A., et al. 2005, 
       Science accepted, astro-ph/0506130 %xray flares
       
 %   \bibitem[1989]{Cardelli} Cardelli, J.~A., Clayton, G.~C. \& Mathis, J.~S. 1989,
 %      ApJ, 345, 245
           
   \bibitem[2005]{GCNMIKE1} Chen, H.-W., Thompson, I., Prochaska, J.~X., et al. 2005,
       GCN Circ. 3709  %MIKE spectrum 1st GCN

   \bibitem[2005]{chenpaper} Chen, H.-W., Prochaska, J.~X., Bloom, J. S., Thompson, I. B. 2005,
       ApJL submitted, astro-ph/0508270  %Chen paper subm.

   \bibitem[2004]{christensen} Christensen, L., Hjorth, J., \& Gorosabel, J. 2004, A\&A, 425, 913

   \bibitem[2005]{GCNSMARTS} Cobb, B.~E., \& Bailyn, C.~D., et al. 2005,
       GCN Circ. 3708  %SMARTS data
   
   \bibitem[2005]{HSThosts} Conselice, C.~J., Vreeswijk, P.~M., Fruchter, A.~S., et al. 2005,
       ApJ preprint doi:10.1086/'432829'
   
%   \bibitem[2005]{GCNOHP} Damerdji, Y., Klotz, A., Boer M. et al. 2005,
%       GCN Circ. 3741 %OHP R band datapoints

   \bibitem[1990]{dickey} Dickey, J.~M., \& Lockman, F.~J. 1990, ARA\&A, 28, 215
       
%   \bibitem[2005]{GCNXMM} Ehle, M. \& Juarez, B. 2005,
%       GCN Circ. 3713 %XMM Newton quick-look     

%   \bibitem[2005]{GCNUVES1} D'Elia, V., Melandri, A., Fiore, F., et al 2005,
%       GCN Circ. 373746 %UVES and FORS1 
       
    \bibitem[2005]{ellisonhalllira}
         Ellison, S. L., Hall, P. B., Lira, P. 2005, AJ in press,
	 astro-ph/0507418

%    \bibitem[2005]{erb}
%        Erb, D.~K., Shapley, A.~E., Pettini, M., et al., in preparation

     \bibitem[2005]{fiore}
        Fiore, F.,  D'Elia, V., Lazzati, D., et al. 2005, ApJ, 624, 853
  
   \bibitem[2005]{hst} Fruchter, A., et al. 2005, ApJ submitted

%   \bibitem[2003]{Fynboz} Fynbo, J. P. U., Jakobsson, P., M{\o}ller, P., et al. 2003, A\&A, 406, L63 %Lya section, metallicity of hosts (to replace Jakbosson et al 05)

%    \bibitem[2005]{GRB050401_DLA} Fynbo, J.~P.~U., Jensen, B.~L., Hjorth, J., et al. 2005,
%    GCN Circ. 3176 % DLA, can't remember N_H value. 
      
%   \bibitem[2005]{GRB050319_probably} Fynbo, J.~P.~U., Hjorth, J., Jensen, B.~L., et al. 2005
%    GCN Circ. 3136 % In GCN: "We find several absorption features, including strong 
    %Lyman-alpha, OI+SiII, SiIV and CIV, corresponding to a redshift of z=3.24. 
  
   \bibitem[2002]{gar02} Garnett, D. R. 2002, 
      ApJ, 581, 1019
	
%   \bibitem[2005]{GCNLT} Gomboc, A., Guidorzi, C., Steele, I.~A., et al. 2005,
%       GCN Circ. 3706 % Liverpool Telescope OT
       
   \bibitem[1998]{gs98} Grevesse, N., \& Sauval, A.J. 1998, 
       Space Sci Rev, 85, 161     
   \bibitem[2005]{GCNXRT} Grupe, D., Kennea, J.~A., \& Burrows, D.~N. 2005,
       GCN Circ. 3714 %XRT, X-ray flares reported
       
   \bibitem[2005]{GCNPROMPT} Haislip, J., Kirschbrown, J., Reichart, D., et al. 2005,
       GCN Circ. 3712 %PROMPT data

   \bibitem[2003]{GRB020124_DLA} Hjorth, J., M{\o}ller, P., Gorosabel, J., et al. 2003,     
    ApJ, 597, 699 %DLA, log(N_H) = 21.7, metal lines
    
   \bibitem[2005]{GCNSwiftBat1} Holland, S.~T., Barthelmy, S., Burrows, D.~N., et al. 2005,
       GCN Circ. 3704 %Swift GCN
           
   \bibitem[2005]{GCNHolman}  Holman, M., Garnavich, P., \& Stanek, K.~Z. 2005,     
       GCN Circ. 3727 % corrected photometry and break   
       
%    \bibitem[2005]{GCNMagellan} Holman, M., Garnavich, P. \& Stanek, K.~Z. 2005,
%       GCN Circ. 3716 %Magellan IMACS spectrum   
          
%   \bibitem[2005]{GCNamateur} Jacques, C. \& Pimentel, E. 2005,  
%       GCN Circ. 3711 % Amateur data
       
   \bibitem[2001]{GRB000301C_DLA} Jensen, B.~L., Fynbo, J.~U., Gorosabel, J. et al. 2001,
    A\&A, 370, 909 % GRB000301C, has a DLA with log(N_H) approx 21.2
 
    \bibitem[2005]{UVLya} Jakobsson, P., Bjornsson, G., Fynbo, J.~P.~U., et al. 2005, MNRAS in press, astro-ph/0505542

%   \bibitem[2004]{GRB030429_DLA} Jakobsson, P., Hjorth, J., Fynbo, J.~P.~U., et al. 2004,
%    A\&A, 427, 785 % DLA log(N_H) = 21.6, metal lines
    
%   \bibitem[2005]{kewleyellison} Kewley, L. J., \& Ellison, S. L., 
%    in preparation    
   
   \bibitem[2005]{KingXrayflares}  King, A., O'Brien, P.~T., Goad, M.~R., et al. 2005,
       ApJL accepted, astro-ph/0508126 %X-ray flare theory
       
    \bibitem[2004]{KK04} 
        Kobulnicky, H. A., \& Kewley, L. J. 2004, ApJ, 617, 240

    \bibitem[2003]{k03}
        Kobulnicky, H. A., Willmer, C. N. A., Weiner, B. J., et al. 2003, 
	ApJ, 599, 1006   
   
%    \bibitem[1998]{GRB971214_no_DLA} Kulkarni, S.~R., Djorgovski, S.~G., Ramaprakash, A.~N., et al. 1998   
%    Nat, 393, 35  %Ly alpha emission, no DLA seen (HG spectrum)
    
%   \bibitem[2005]{GCNTAROT} Klotz, A., Boer M. \& Atteia, J.L. 2005,
%       GCN Circ. 3720  % TAROT early lightcurve.
       
    \bibitem[2004]{lam04}
        Lamareille, F., Mouhcine, M., Lewis, I., et al. 2004, 
	MNRAS, 350, 396
   
%   \bibitem[2005]{GCNSwiftBat2} Markwardt, C.~B., Barbier, L., Barthelmy S., et al. 2005,
%       GCN Circ. 3715 %Revised BAT analysis
    \bibitem[2004]{mollerLZ} M{\o}ller, P., Fynbo, J. P. U., \& Fall, S. M. 2004,
         A\&A, 422, L33 
    	
%    \bibitem[2002]{Moeller021004} M{\o}ller, P., Fynbo, J. P. U., Hjorth, J., et al. 2002,
%         A\&A, 396, 21 
	
   \bibitem[2004]{ml04} Murphy, M. T., \& Liske, J. 2004, 
	MNRAS, 345, L31
   
%   \bibitem[2005]{GCNXRTrefined} Perri, M., Capalbi, M.,  Giommi, P. et al. 2005,
%       GCN Circ. 3722  %XRT refined analysis. Columns.
       
   \bibitem[1992]{pei92} Pei, Y. 1992, 
	ApJ, 395, 130    
   
%   \bibitem[2005]{GCNMIKE2} Prochaska, J.~X., Chen, H.-W., Bloom, J.~S., et al. 2005,
%       GCN Circ. 3732  %MIKE spectrum 2nd GCN
       
%   \bibitem[2005]{GRB050502a_maybe} Prochaska, J.~X., Ellison, S., Foley, R.~J., et al. 2005
%    GCN Circ. 3332 % broad Ly alpha absorption. Sara?   
   
   \bibitem[2005]{GCNWHT} Rol, E., Starling, R.~L.~C., Wiersema, K., et al. 2005,
       GCN Circ. 3710 % Our spectrum

%   \bibitem[2004]{Savaglio} Savaglio, S., \& Fall, S. M. 2004, ApJ, 614, 293

    \bibitem[2003]{Schaefer}
    Schaefer, B. E., Gerardy, C. L., Höflich, P., et al. 2003, ApJ, 588, 387
   
   \bibitem[1998]{Schlegel98} Schlegel, D.~J., Finkbeiner, D.~P., \& Davis, M. 2004,
       ApJ, 500, 525
   \bibitem[2004]{alice04} Shapley, A. E., Erb, D. K., Pettini, M., et al. 2004, 
	ApJ, 612, 108
	
%	\bibitem[2002]{Silva} Silva, A. I., \& Viegas, S. M. 2002, MNRAS, 329, 135
        
   \bibitem[2005]{GCNOT} Sota, A., Castro-Tirado, A.~J., Guziy, S., et al. 2005,
       GCN Circ. 3705  %Discovery OT
      
   \bibitem[2005]{Starling_021004} Starling, R. L. C., Wijers, R. A. M. J., Hughes, M. A., et al. 2005,
        MNRAS, 360, 305

   \bibitem[2004]{nial} Tanvir, N. R., Barnard, V. E., Blain, A. W., et al. 2004, MNRAS, 352, 1073 	 
       
    \bibitem[2004]{tremonti}  Tremonti, C.,  Heckman, T. M., Kauffmann, G., et al. 2004, 
    ApJ, 613, 898 
   
%   \bibitem[2001] VVan Dokkum, P. 2001,
%     PASP 113, 1420 
     
 \bibitem[2004]{GRB030323_DLA} Vreeswijk, P.~M., Ellison, S.~L., Ledoux, C., et al. 2004,    
    A\&A, 419, 927 %DLA, log(N_H) = 21.9
  
   \bibitem[2000]{wabs} Wilms, J., Allen, A., \& McCray, R. 2000, ApJ, 542, 914 

% \bibitem[2005]{GRB011211_DLA} Vreeswijk, P.~M, Smette, A., Fruchter, A.~S. et al. 2005,
%    A\&A, in prep %DLA with log(N_H) = 20.4, smal extinction, 3 metallicities: Si,Al and Fe
\end{thebibliography}
\end{document}